\documentclass[submitted]{IEEEtran} 
\usepackage{fancyhdr}
\usepackage{enumerate}
\usepackage{graphicx}
\usepackage{url}
\usepackage{cite}
\usepackage{color}
\usepackage{upgreek}
\usepackage{longtable}
\usepackage[colorlinks=true,linkcolor=blue,urlcolor=blue,citecolor=blue]{hyperref}
\usepackage[colorinlistoftodos]{todonotes}

\usepackage{changes} 

\begin{document}
	
	\title{\bf A Determination of the Local Gravitational Acceleration for the Tsinghua \\ Tabletop Kibble Balance}
	\author{Weibo Liu, Nanjia Li, Yongchao Ma, Ruo Hu, Shuqing Wu, Wei Zhao, Songling Huang, \\
		\textit{Senior Member, IEEE}, Shisong Li$^{\dagger}$, \textit{Senior Member, IEEE}
		
		\thanks{W. Liu, N. Li, Y. Ma, W. Zhao, S. Huang, and S. Li are with the Department of Electrical Engineering, Tsinghua University, Beijing 100084, China.}
		\thanks{R. Hu and S. Wu are with the National Institute of Metrology (NIM), Beijing 100029, China. }
		\thanks{This work was supported by the National Key Research
			and Development Program under Grant 2022YFF0708600 and the National Natural Science Foundation of China under Grant 52377011.}
		\thanks{$^\dagger$Email: shisongli@tsinghua.edu.cn}}
	
	\maketitle
	
	\begin{abstract}
		The Kibble balance requires a measurement of the local gravitational acceleration, $g$, with a typical {relative measurement} uncertainty of $10^{-9}$. In this paper, the determination of $g$ for the Tsinghua tabletop Kibble balance is presented. A polynomial fitting method is proposed for blind transfers of the absolute gravitational acceleration using relative gravimeters, showing agreement with the value obtained by the tide correction within a few parts in $10^{9}$. Horizontal and vertical gravity gradients are extracted by mapping the gravity distribution at different heights. The self-attraction effect of major components in the experiment, as well as some time-varying systematic effects, are modeled. The final determination of the gravitational acceleration at the {mass position}, with an uncertainty of 5.4\,$\upmu$Gal ($k=2$), is achieved for the Tsinghua tabletop Kibble balance experiment. 
	\end{abstract}
	
	
	\section*{Nomenclature}
	\begin{itemize}
		\item  [AG] - Absolute gravimeter
		\item  [RG]  - Relative gravimeter
		\item  [NIM]  - National Institute of Metrology, China
		\item  [THU]  - Tsinghua University
		\item  [SG]  - Superconducting gravimeter
		\item  [HGG]  - Horizontal gravity gradient
		\item  [VGG]  - Vertical gravity gradient
		\item  [FEA]  - Finite Element Analysis
		\item  [IERS] - International Earth Rotation and Reference System Service
		
		\item  [$\Delta g_{\rm{s}}$]  - Gravity correction of instrument self-attraction
		\item  [$\Delta g_{\rm{t}}$]  - Time-varying gravity correction
		\item  [$\Delta g_{\rm{ta}}$]  - Gravity correction of atmospheric mass effect
		\item  [$\Delta g_{\rm{tp}}$]  - Gravity correction due to the polar motion
		\item  [$\Delta g_{\rm{tt}}$]  - Gravity correction of the Earth's tides
		\item  [$\Delta g_{\rm{HGG}}$] - Gravity correction related to horizontal field distribution
		\item  [$\Delta g_{\rm{VGG}}$] - Gravity correction related to vertical field distribution
		\item  [$g_{\rm{r}}$]  - Absolute gravitational acceleration at the reference site
		\item  [$g_{\rm{0}}$]  - Absolute gravitational acceleration at the reference point of the Kibble balance site
		\item  [$g_{\rm{m}}$]  - Absolute gravitational acceleration at the {mass position} of the Kibble balance site
		\item  [$g$] - Absolute local gravitational acceleration
		\item  [$m$]  - Test mass to be calibrated by Kibble balance
		\item  [$Bl$]  - Magnetic geometrical factor of a Kibble balance
		\item  [$I$]  - Current through the coil in the weighing phase
		\item  [$U$]  - Coil induced voltage in the velocity phase
		\item  [$v$]  - Coil moving velocity in the velocity phase
		\item  [$O$]  - Polynomial fit order
		\item  [$\tau$] - Time delay of tides in two sites
		\item  [$\Delta g$] - Gravity offset of two sites/measurements
		\item  [$H$] - Height of the horizontal gravity mapping
		\item  [$h_0$] - Height of the CG6 sensor to the ground surface
		\item  [$H_{\rm{lab}}$] - Height of the Kibble balance laboratory
		\item  [$p$] - Pressure (in hPa) at the {mass position}
		\item  [$\omega$] - Angular velocity of the Earth’s rotation
		\item  [$R$] - Radius of the Earth
		\item  [$\lambda,\phi$] - Geodetic coordinates of the measurement point
		\item  [$x,y$] - Pole coordinates updated by the IERS
		\item  [$k_{\rm{op}}$] - Gain factor of the Tsoft tide estimation
		\item  [$\Delta g_{\rm{o}}$] - Offset of the Tsoft tide estimation
		\item  [$z_{\rm{w}}$] - Distance between the bottom of the vacuum chamber and the {mass position}

	\end{itemize}

	\section{Introduction}
	
	\IEEEPARstart{T}{he Kibble} balance, an instrument originally proposed by Dr. Bran Kibble \cite{Kibble1976} offering a precision link between a test mass $m$ and the Planck constant $h$, stands as one of the primary methods for mass realizations at the kilogram level and beyond. Presently, numerous metrology institutes are actively engaged in Kibble balance experiments, exemplified by initiatives from institutions such as \cite{NRC, NIST, NIST2, METAS, LNE, BIPM, MSL, NIM, KRISS, UME, PTB, NPL}.
	
	
	A conventional Kibble balance comprises two distinctive measurement phases: the weighing phase and the velocity phase. In the weighing phase, the gravitational force acting on the test mass, $mg$, is counterbalanced by an electromagnetic force generated by a current-carrying coil in a magnetic field, expressed as $mg=BlI$, where $I$ denotes the current in the coil and $Bl$ represents the dot product of magnetic flux density and current flowing path, so-called the magnetic geometrical factor. Subsequently, in the velocity phase, the $Bl$ term is calibrated by moving the coil with a velocity $v$ in the same magnetic field, resulting in an induced voltage on coil terminals $U=Blv$. With $Bl$ eliminated, the mass is determined as $m=\frac{UI}{gv}$. Details of the Kibble balance principle can be found in recent review papers, e.g.~\cite{Stephan16}. The most accurate Kibble balance can calibrate mass $m$ with a {relative} uncertainty of about one part in $10^8$ \cite{NRC}, hence necessitating a determination of the local gravitational acceleration $g$ at the order of $10^{-9}$, e.g.~\cite{liard2014gravimetry,jiang2013gravimetric,choi2017gravity,merlet2008micro,leaman2015determination,louchet2011comparison,xu2020determination,li2017self,METAS}.
	
	In late 2022, Tsinghua University initiated a tabletop Kibble balance project with the objective of creating an accurate, robust, compact, and open-hardware mass realization instrument \cite{li2022design,li2023design}. {The goal of the measurement uncertainty is below 50$\,\upmu$g for mass calibrations ranging from 10\,g to 1\,kg.} Progress in the Tsinghua tabletop Kibble balance measurement system has recently transpired. Herein, we report the determination of the local gravitational acceleration, $g$, for the Tsinghua tabletop Kibble balance. A generalized procedure for $g$ measurement is presented, accompanied by a detailed discussion of the measurement specifics.
	
	Note that to be consistent with the gravity community, the non-SI unit, Gal, is used in the following manuscript, where 1\,Gal$=1\times10^{-2}\,$m/s$^{2}\approx 10^{-3}\, g$, 1\,mGal$=1\times10^{-5}\,$m/s$^{2}\approx 10^{-6}\,g$ and $1\,\upmu$Gal$=1\times10^{-8}$\,m/s$^2$$\approx 1\times 10^{-9} \, g$.
	
	
	

	\section{Generalized $g$ measurement principles}
	
\begin{figure}[tp!]
		\centering
		\includegraphics[width=0.5\textwidth]{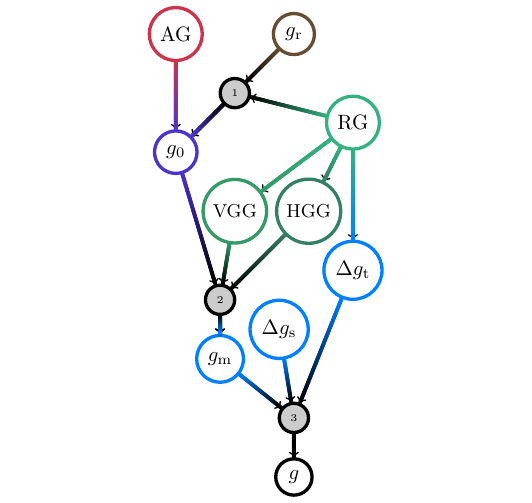}
		\caption{A schematic overview for the absolute determination of gravitational acceleration ($g$) in Kibble balances is presented. In the diagram, AG and RG represent absolute and relative gravimeters, respectively. $g_{\rm{r}}$ denotes the value at an external reference site, while $g_{\rm{0}}$ represents the value at the reference site within the Kibble balance room. VGG and HGG depict the vertical and horizontal gravity gradients, respectively. $g_{\rm{m}}$ signifies the transferred value at the {mass position}. Corrections, primarily related to tides, are encapsulated in $\Delta g_{\rm{t}}$, and $\Delta g_{\rm{s}}$ accounts for the correction stemming from the instrument's self-attraction. The algorithms for $g$ calculations are denoted as 1, 2, and 3.
		}
		\label{fig:scheme}
	\end{figure}

	Fig. \ref{fig:scheme} illustrates a generalized schematic flow for determining the local gravitational acceleration ($g$) in Kibble balance experiments. The process involves three key steps, elaborated upon below.
	
	\begin{enumerate}
		\item \textit{Determine the absolute gravitational acceleration $g_{\rm{0}}$ at a reference position in the Kibble balance site.}
		
		In this step, two general methods are typically employed. The first involves the setup of an absolute gravimeter, such as classic free-falling interferometry~\cite{niebauer1995new} or an atomic gravimeter~\cite{peters1999measurement}, at the reference position of the Kibble balance room. This allows for the direct measurement of the absolute value of $g_{\rm{0}}$. The second method entails transferring the absolute gravitational acceleration value, $g_{\rm{r}}$, from a reference site (absolute $g$ value is precisely known) using relative gravimeters. In this approach, it is crucial to account for time-varying corrections during the measurement, particularly those related to the Earth's tides. 
		{Typically, repeated ABA or ABBA measurements are required to reduce the Type A uncertainty and the Type B uncertainty components caused by the instrument drift. Performing more measurement loops in a short period of time helps to reduce both types of uncertainties, and the number of measurements at each site, $N$ depends on the best overall uncertainty required. For example, if a CG6 relative gravimeter (standard deviation $<$5\,$\upmu$Gal, compensated daily drift $<$20\,$\upmu$Gal) is used, $N$ greater than 6 can result in an instrument-related uncertainty of less than 2\,$\upmu$Gal. }
	
	
	\item \textit{Determine the absolute gravitational acceleration $g_{\rm{m}}$ at {the mass position} in Kibble balances.}
	
	In Kibble balances, the gravitational acceleration at the mass location, $g_{\rm{m}}$, is required. Therefore, a gravity transfer from the reference position of the Kibble balance to {the mass position} ($g_{\rm{0}}\rightarrow g_{\rm{m}}$) should be performed. For this purpose, a mapping of the gravity gradient in both vertical and horizontal directions (VGG and HGG) must be established. The conventional approach is to measure the gravity change at various known horizontal and vertical positions using relative gravimeters and then apply a fitting algorithm to determine the difference of the gravitational acceleration, $g_{\rm{m}}-g_{\rm{0}}$.
	
	\item \textit{Determine the final gravitational acceleration $g$ considering time-varying and self-attraction corrections.}
	
	The main errors in the gravitational acceleration $g$ obtained by the above steps include $\Delta g_{\rm{t}}$, time-varying corrections, mainly the Earth’ tides, and $\Delta g_{\rm{s}}$, the correction from the self-attraction of the instrument. The tides lead to corrections depending on the time of day, the latitude, the longitude, and the altitude. The tide correction can be measured by a relative gravimeter or using theoretical estimation tools such as Tsoft~\cite{van2005tsoft}, ETERNA~\cite{wenzel1996nanogal}, Tamura tidal potential~\cite{hartmann1995hw95}, etc. The instrument's self-attraction can be corrected by direction measurement or finite element analysis (FEA) calculations following the similarity between the law of universal gravitation and Coulomb's law.
\end{enumerate}

In the following sections, the determination of $g$ for the Tsinghua tabletop Kibble balance following the above three steps is presented. 

\section{$g_{\rm{0}}$ Determination}
\label{sec02}

\begin{figure}[tp!]
	\centering
	\includegraphics[width=0.5\textwidth]{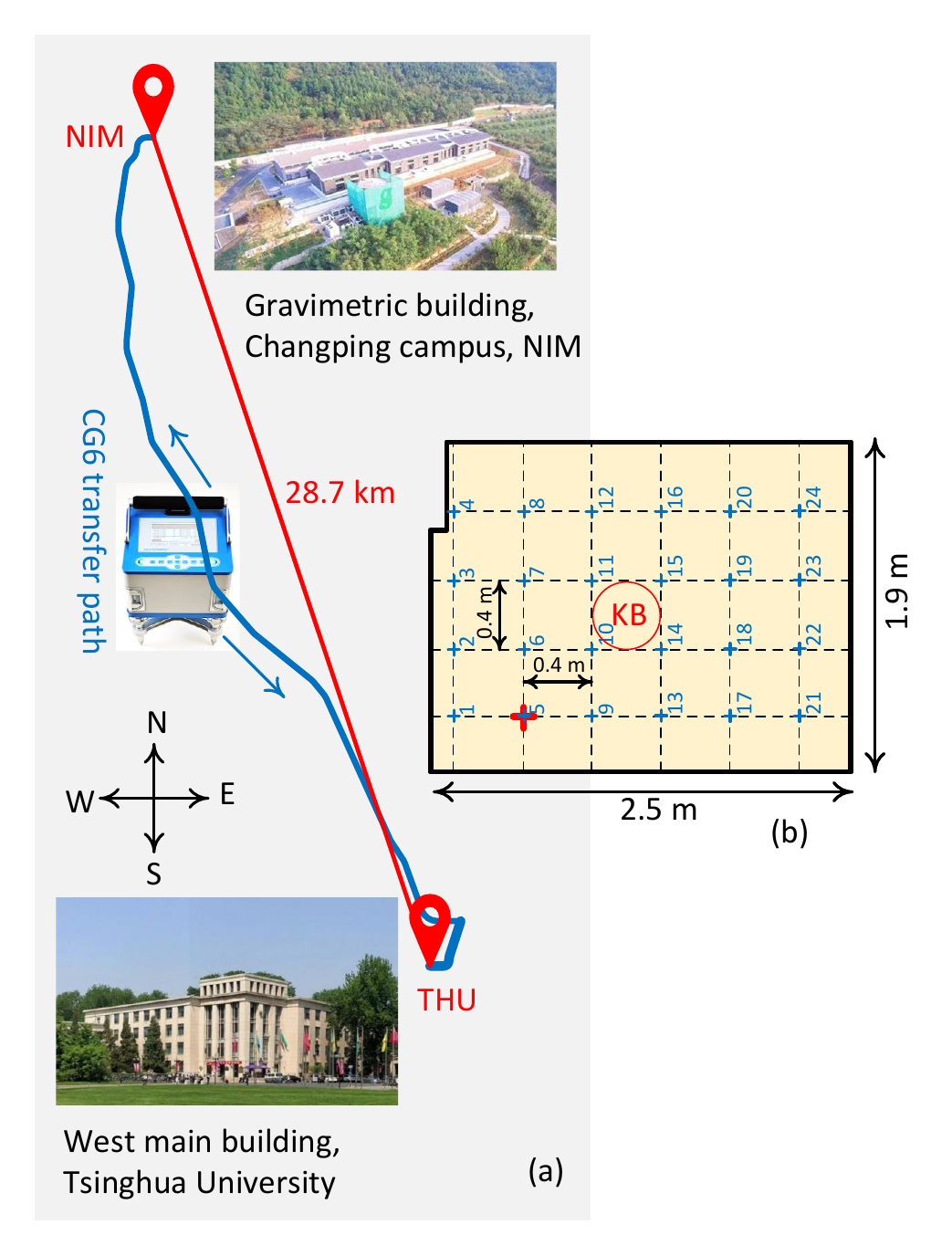}
	\caption{(a) describes the geographical positions of the NIM and THU laboratories, accompanied by the transmission path of the CG6 throughout the measurements. (b) demonstrates the arrangement of 24-point locations in the Tsinghua lab, where the red points serve as the reference point.}
	\label{fig:KBsite}
\end{figure}

For the Tsinghua Kibble balance, the relative gravity transfer approach ($g_\mathrm{r}\rightarrow g_0$) is employed to determine $g_{\rm{0}}$ at the Kibble balance site. In Fig. \ref{fig:KBsite} (a), the reference site providing the $g_\mathrm{r}$ value is a point utilized in the 2017 Key Comparison of Absolute Gravimeters (CCM.G-K2.2017) \cite{stock2023final}, situated at the Gravimetric building, Changping campus of the National Institute of Metrology (NIM, China). The $g_{\rm{r}}$ value is recorded as 980\,122\,922.8\ $\upmu$Gal with a standard uncertainty of 1.0\ $\upmu$Gal.


The Tsinghua tabletop Kibble balance is situated in a laboratory on the fourth floor of the west main building, Tsinghua University. This distinctive Kibble balance is installed within a building rather than on the Earth's surface or in an underground base, with the purpose of testing its resilience in a general-conditioned laboratory environment. Fig. \ref{fig:KBsite} (b) depicts the top view of the measurement site, with the Tsinghua tabletop Kibble balance anticipated to be installed near the center of the 2.5\,m\,$\times$\,1.9\,m room. A gravity reference point, denoted by the red cross in Fig. \ref{fig:KBsite}(b), is chosen as the $g$ transferred point. Positioned approximately 0.85\,m away from the vacuum chamber, this allows space for future absolute $g$ checks or relative gravity monitoring.

As illustrated in Fig. \ref{fig:KBsite} (a), the linear distance between the reference gravity site (NIM) and the Tsinghua Kibble balance site (THU) is approximately 28.7\,km. To measure the gravity difference between NIM and THU sites, a relative gravimeter {CG6 (Scintrex)~\cite{CG6}, a spring-type relative gravimeter with a measurement range of 8000\,mGal, resolution of 0.1\,$\upmu$Gal, standard deviation $<$5\,$\upmu$Gal, and compensated daily drift $<$20\,$\upmu$Gal,} is utilized. 
Following the transportation route (blue line, approximately 33\,km drive), the CG6 is transferred between the two sites using a vehicle, alternately measuring the gravitational acceleration. The measurement, lasting about 14 hours, involves seven loops. Implementing the ABA measurement strategy allows the removal of the linear drift of the instrument, enabling precise determination of the gravity difference.

\begin{figure}[tp!]
	\centering
	\includegraphics[width=0.5\textwidth]{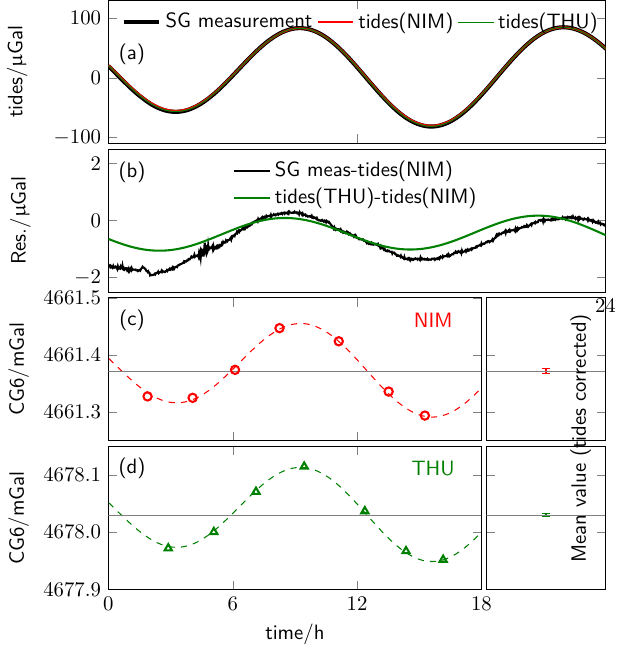}
	\caption{(a) illustrates the tidal estimations at the NIM and THU sites, alongside the relative $g$ measurements at the NIM site using the superconducting gravimeter {iGrav-SG}. In (b), the disparities between the tide estimations and measurements are presented. (c) and (d) showcase the CG6 measurement outcomes at NIM and THU, respectively. The dashed lines represent the tides estimated by Tsoft. The right side of (c) and (d) displays the average gravity values after the tide correction.}
	\label{fig:tides}
\end{figure}

Given the geographical separation of the two measurement sites, variations in gravity corrections due to the Earth's tides may impact the measurement results differently. In consideration of this discrepancy and to streamline the measurement process, the tide correction within the CG6 is disabled. To account for tidal influences, Tsoft~\cite{van2005tsoft}, a program incorporating date, latitude, longitude, and altitude inputs to calculate tidal effects on gravity, is employed. Independent calculations of tidal influence, presented in Fig. \ref{fig:tides} (a) and (b), are conducted for NIM and Tsinghua, respectively.

As illustrated, there is a difference of approximately -0.5\,$\upmu$Gal in the mean value and a 1\,$\upmu$Gal peak-to-peak value between the tidal influences, attributed to a phase shift (time delay). To validate the accuracy of the tide estimation {by Tsoft}, an experimental measurement using a superconducting relative gravimeter {iGrav-SG (GWR Instruments)} at the NIM site\cite{xu2020determination} was conducted during the $g$ transfer period. {The iGrav-SG~\cite{Warburton2010InitialRW} has a precision of $10^{-3}\,\upmu$Gal in the frequency domain and 0.05\,$\upmu$Gal in the time domain for one-minute averaging, and can be used as an ultra-high precision continuous gravity reference for measurements.} As depicted in Fig. \ref{fig:tides} (b), the maximum difference between the experimental and Tsoft estimation is within 2\,$\upmu$Gal.

The seven-loop CG6 measurement results at the NIM and THU sites are presented in Fig. \ref{fig:tides} (c) and (d). Following the tide removal process, the mean $g$ measurement at the NIM site is 4\,661.371\,9\,mGal, with a standard deviation of 4.2\,$\upmu$Gal. Similarly, at the THU Kibble balance site, the mean measurement is 4\,678.030\,3\,mGal, with a standard deviation of 3.0\,$\upmu$Gal. The slightly lower data deviation at the THU site indicates {that the vibration noise floor of the Tsinghua Kibble balance laboratory may be low, at least in the sensitive range of the CG6}. This could be attributed to the proximity of the laboratory to the building mass center. The gravity difference between the THU site and the NIM site is calculated as 16.658\,4\,mGal, with a standard uncertainty of 2.1\,$\upmu$Gal. Using the NIM reference value, the gravitational acceleration at the reference point in the Tsinghua tabletop Kibble Balance room is determined as $g_{\rm{0}}=(980\,139\,581.2\pm2.3) \, \upmu$Gal ($k=1$).

\begin{figure}[tp!]
	\centering
	\includegraphics[width=0.5\textwidth]{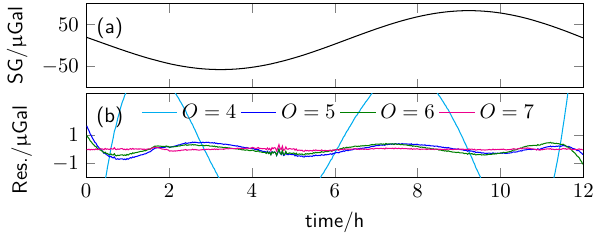}
	\caption{(a) presents a 12-hour tide measurement result using the superconducting gravimeter {iGrav-SG}. (b) shows the differences between the experimental and the $O$-order polynomial fittings. The fitting orders, $O=4$ to $O=7$, are shown in the plot.}
	\label{fig:fit}
\end{figure}

During the data processing, we observed that if the tide difference between the reference site (NIM) and the measurement site (THU) is negligible, a polynomial fit method can estimate the gravity difference without prior knowledge of tide information. The concept is to assume that over a short period, such as 12 hours, the tide correction's shape can be adequately represented by a polynomial fit of order $O$, e.g., $O=6$. This assumption can be validated using data from the superconducting gravimeter {iGrav-SG}. Fig. \ref{fig:fit} displays the residuals of a 12-hour tide measurement obtained by the superconducting gravimeter {iGrav-SG} and polynomial fittings of various orders ($O=4,5,6,7$). Notably, for $O\ge5$, the tide difference between the experimental data and the fit remains within $\pm1 \ \upmu$Gal. 

With a fixed-order polynomial fit representing the tides, the $g$ transfer can be simplified as
\begin{eqnarray}
	&&\textbf{Min}\{\textbf{Res.}\{\mathrm{Polyfit}\{g_1(t), g_2(t-\tau)+\Delta g\}\}\}\nonumber\\
	&&\textbf{s.t.}~\tau=\mathrm{const.}, \Delta g=\mathrm{const}.
\end{eqnarray}
where $g_1(t)$ and $g_2(t)$ are the time series of the relative $g$ measurement at two sites, $\tau$ (time delay of tides in two sites) and $\Delta g$ (gravity offset) constant parameters to be determined. The goal of the estimation is to make the lowest residual of the polynomial fit. 

\begin{figure}[tp!]
	\centering
	\includegraphics[width=0.5\textwidth]{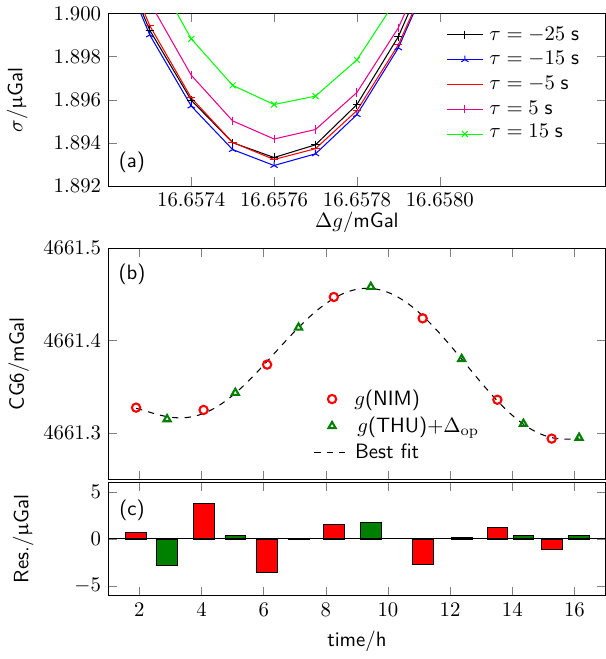}
	\caption{(a) plots the standard deviation of the residual for the polynomial fit at different gravity offsets $\Delta g$ and time delay $\tau$.  (b) illustrates the best polynomial fit curve and data points, while (c) presents the residuals of the best fit.}
	\label{fig:polyfit}
\end{figure}

Here we take the CG6 data as an example and set the fit order $O=6$. $g_1$ and $g_2$ are respectively the measurement results at the NIM and THU sites. The standard deviation of the residual, $\sigma$, is set as the target for minimization. As shown in Fig.\ref{fig:polyfit}, by scanning $\tau$ and $\Delta g$, the optimal parameters, $\tau_\mathrm{op}=15$\,s and $\Delta g_\mathrm{op}=16.657\,6$\,mGal, are obtained. In this case, the standard deviations of the NIM measurement and THU measurement are respectively 2.5\,$\upmu$Gal and 1.5\,$\upmu$Gal. This blind determination of the gravitational acceleration at the reference point in the Tsinghua tabletop Kibble Balance room yields $g_0=(980\,139\,580.4\pm1.6) \, \upmu$Gal ($k=1$). The result agrees well with the value obtained by the conventional tide correction and has a slightly lower measurement uncertainty. 


\section{$g_{\rm{m}}$ Determination}
\label{sec03}

The preceding section establishes the absolute gravity value at a designated reference point within the Tsinghua Kibble balance room. In order to ascertain the gravity value at {the specific experimental mass position}, located approximately 0.85 m horizontally and 1 m vertically from the reference point on the floor, it is imperative to determine both the horizontal gravity gradient (HGG) and vertical gravity gradient (VGG). This necessitates the mapping of the gravity distribution across the experiment room.

\begin{figure}[tp!]
	\centering
	\includegraphics[width=0.5\textwidth]{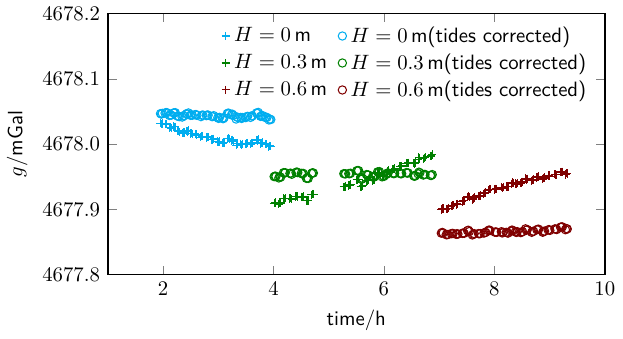}
	\caption{Relative gravity measurement results at different heights ($H=0$\,m, 0.3\,m, 0.6\,m). For each group of data, from left to right correspond points 1 to 24. }
	\label{fig:HGG1}
\end{figure}

As presented in Fig. \ref{fig:KBsite} (b), the floor of the Tsinghua Kibble balance laboratory is segmented into a grid arrangement of 24 points, organized in four rows and six columns, labeled from 1 to 24, with point 5 serving as the reference position. The gravitational acceleration ($g$) values at three distinct height levels, i.e. $H=0$\,m, $H=0.3$\,m, and $H=0.6$\,m, are systematically mapped using the CG6 instrument. The outcomes of these measurements, segregated by height levels, are presented in Fig. \ref{fig:HGG1}. By subtracting the tidal influence from the measured data, it becomes feasible to derive the relative gravity distribution in three-dimensional space.

\subsection{HGG}

\begin{figure}[tp!]
	\centering
	\includegraphics[width=0.475\textwidth]{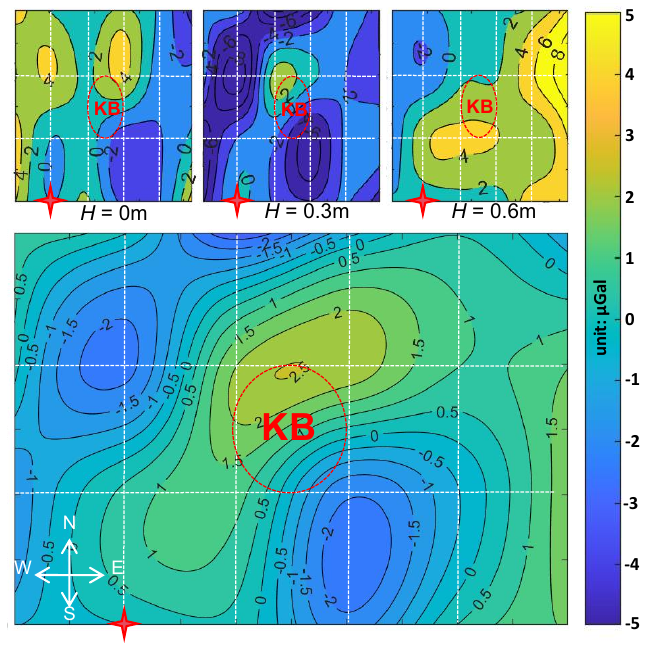}
	\caption{The upper subplots show, from left to right, the change in g at ground level, 0.3\,m above ground level, and 0.6\,m above ground level. The unit of the contour is $\upmu$Gal. The lower subplot is an average of the top three plots.  The red circle, KB, is where the Tsinghua Kibble balance will be installed. }
	\label{fig:HGG2}
\end{figure}

The Kibble balance room is situated in the central area of the west main building at Tsinghua University, characterized by a flat terrain with an altitude of approximately 64\,m above sea level on the ground floor. Given the relatively confined dimensions of the Kibble balance experimental space (1.9\,m$\times$2.5\,m), it is anticipated that the horizontal gradient (HGG) is insignificantly small. The top subplots of Fig. \ref{fig:HGG2} depict the relative gravitational distribution at three distinct heights, $H=0$\,m, 0.3\,m, and 0.6\,m. Note that the relative $g$ value at the reference position (point 5) is normalized to zero, and cubic interpolation is employed to enhance the smoothness of the plots. 
The lower subplot represents the average of the three plots. The designated location for the installation of the Tsinghua Kibble balance is denoted by the red circle labeled 'KB'.

Despite inherent measurement uncertainties, it is challenging to discern a distinct horizontal gravity gradient along the horizontal axes. To simplify the analysis, the mean of the mapping within the central rectangle (formed by points 10-11-15-14) is employed to quantify the $g$ difference between the reference point and {the experimental mass position}. This yields differences of $(1.2\pm1.5)\,\upmu$Gal, $(-0.7\pm2.7)\,\upmu$Gal, and $(3.2\pm0.9)\,\upmu$Gal for $H=0$\,m, $H=0.3$\,m, and $H=0.6$\,m, respectively. The weighted mean of the three measurements is $(2.4\pm0.7)\,\upmu$Gal. A cross-check directly using the measurement data at points 10, 11, 14, and 15 is conducted, yielding an average of $(2.2\pm1.2)\,\upmu$Gal, which agrees well with the value obtained by fittings. Notably, no substantial variation in HGG is observed as a function of height ($H$).

To further scrutinize the HGG, the gravity maps at the three distinct levels are averaged and presented in the lower subplot of Fig. \ref{fig:HGG2}. It can be seen that the gravity distribution remains within a few $\upmu$Gal, with no discernible global directional gradient. 


\subsection{VGG}

\begin{figure}[tp!]
	\centering
	\includegraphics[width=0.5\textwidth]{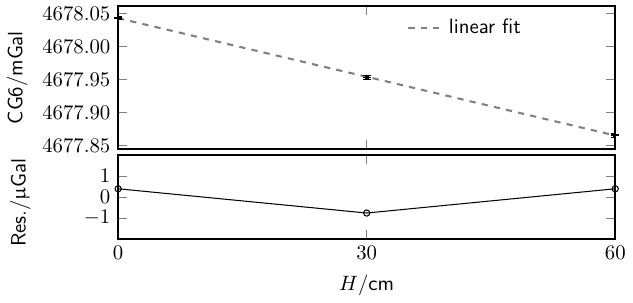}
	\caption{The upper subplot displays the curve fitted for the vertical gravity gradient. The lower subplot depicts the residuals at varying heights.}
	\label{fig:VGG}
\end{figure}

The data presented in Fig. \ref{fig:HGG1} illustrates a linear change in $g$ along the vertical direction. To assess this linearity, linear fits were conducted using three distinct height $g$ data points for each location, and the average of all 24 fitted curves is depicted in Fig. \ref{fig:VGG} (a). The resulting ultimate curve from this fitting process is characterized by $\partial g/\partial H=-296.2\,\upmu$Gal/m, showcasing an exceptionally linear fit with $R^2\approx0.99994$ and residuals below 1\,$\upmu$Gal, as illustrated in Fig. \ref{fig:VGG} (b).

Given that $g$ changes by $-2.962\,\upmu$Gal per centimeter vertically, precise measurement of the vertical distance between {the mass position} and the CG6 sensor at the reference point becomes crucial. Considering the sensor height of the CG6, the $g$ value at {the mass position} can be expressed as
\begin{equation}
	g_{\rm{m}}=g_{\rm{0}}+\Delta g_{\mathrm{HGG}}+\Delta g_{\mathrm{VGG}},
\end{equation}
where $\Delta g_{\mathrm{VGG}}=(\partial g/\partial H)(H-h_0)$; $H$ represents the height of {the mass position} from the floor surface, and $h_0$ denotes the height of the gravity measurement sensor. For the CG6 experimental setup here, $h_0$ includes the height of a supporting platform and is approximately $21.2$\,cm.

\section{Final result of $g$}

\subsection{Time-varying corrections, $\Delta g_{\rm{t}}$}

In our analysis, we consider three significant time-varying corrections: the atmospheric mass effect, the Earth's tides, and the polar motion effect.

The atmospheric mass effect is estimated in accordance with the US Standard Atmosphere 1976~\cite{atmosphere1976us}. The atmospheric pressure $p$ (in hPa) is measured at the laboratory, and the correction for gravitational acceleration is expressed as
\begin{equation}
	\Delta g_{\rm{ta}}=-0.3\left[p-1013.25\left(1-\frac{0.0065H_{\mathrm{lab}}}{288.15}\right)^{5.2559}\right],
\end{equation}
where $H_{\mathrm{lab}}$ is the elevation of the measurement point. For the Tsinghua tabletop Kibble balance, $H_{\mathrm{lab}}$ is approximately 64\,m. The standard uncertainty assigned for the atmospheric mass correction is 0.5\,$\upmu$Gal.

The gravitational field change due to the polar motion can also be corrected using a formula~\cite{jiang2013gravimetric}, i.e.
\begin{equation}
	\Delta g_{\rm{tp}}=1.16\omega^2R\sin2\phi(x\cos\lambda-y\sin\lambda)\times10^8,
\end{equation}
where $\omega$ is the angular velocity of the Earth's rotation, $R$ is the radius of the Earth, $\lambda$ and $\phi$ represent the geodetic coordinates of the measurement point, and $x$, $y$ are the pole coordinates updated by the International Earth Rotation and Reference System Service (IERS). The unit of the $\Delta g_{\mathrm{tp}}$ correction is $\upmu$Gal, and the standard uncertainty assigned to the polar motion correction is 0.1\,$\upmu$Gal.

The Earth's tide correction constitutes the most substantial time-varying component in $g$ measurement, with a maximum tide variation of up to 280\,$\upmu$Gal or approximately $2.8\times10^{-7}$. For the tide correction, Tsoft is employed to estimate the changes in tides over time, and the parameters used are detailed in Tab. \ref{tab:tidepara}.

\begin{table}[tp!]
	\centering
	\caption{Tsoft parameters used for estimation the Earth's tides.}
	\label{tab:tidepara}
	\begin{tabular}{ccccc}
		\hline
		Group name	&	Min freq	&	Max freq	&	Amplitude factor	&	phase shift	\\
		\hline
		DC	&	0.000000	&	0.000001	&	1.00000 	&	0.0000 	\\
		long	&	0.000140	&	0.002427	&	1.15800 	&	0.0000 	\\
		Mf	&	0.002428	&	0.249951	&	1.15738 	&	0.0000 	\\
		Q1	&	0.721500	&	0.906315	&	1.15424 	&	0.0000 	\\
		O1	&	0.921941	&	0.940487	&	1.15423 	&	0.0000 	\\
		P1	&	0.958085	&	0.998028	&	1.14916 	&	0.0000 	\\
		K1	&	0.999853	&	1.003651	&	1.13493 	&	0.0000 	\\
		PSI1	&	1.005329	&	1.005623	&	1.26951 	&	0.0000 	\\
		PHI1	&	1.007595	&	1.011099	&	1.17025 	&	0.0000 	\\
		OO1	&	1.013689	&	1.216397	&	1.15626 	&	0.0000 	\\
		All2	&	1.719381	&	2.182843	&	1.16188 	&	0.0000 	\\
		M3	&	2.753244	&	3.381478	&	1.07352 	&	0.0000 	\\
		M4	&	3.381379	&	4.347615	&	1.03900 	&	0.0000 \\
		\hline
	\end{tabular}
\end{table}

\begin{figure}[t!]
	\centering
	\includegraphics[width=0.5\textwidth]{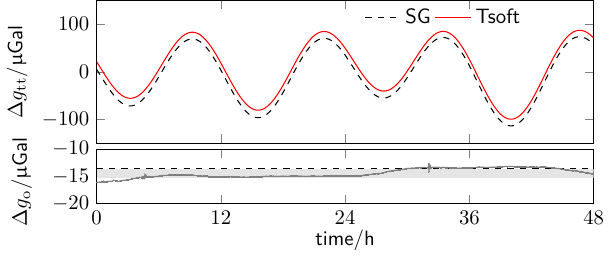}
	\caption{The upper plot shows the tides comparison of a two-day measurement from the superconducting gravimeter {iGrav-SG} and the Tsoft estimation. The lower plot is the difference between two results after the gain $k_{op}$ is multiplied, i.e. $g(\mathrm{SG})-k_\mathrm{op}\cdot g(\mathrm{Tsoft})$. The dashed lines show the boundaries of the standard deviation of the data.}
	\label{fig:Tsoftcal}
\end{figure}

To ensure the accuracy of Tsoft estimation, a calibration procedure is implemented using the superconducting gravimeter {iGrav-SG}. The upper subplot of Fig.\ref{fig:Tsoftcal} illustrates a two-day measurement result from both the superconducting gravimeter {iGrav-SG} and the Tsoft estimation. The calibration aims to determine the optimal gain factor $k$ by minimizing the standard deviation of $g$(SG)-$k\cdot g$(Tsoft). Through numerical search, the optimal gain factor is found to be $k_{\mathrm{op}}=1.008$. The lower subplot of Fig.\ref{fig:Tsoftcal} demonstrates the residual between the superconducting gravimeter {iGrav-SG} measurement and the Tsoft estimation in this case. An offset, $\Delta g_{\mathrm{o}}=(-14.4\pm 0.8)\,\upmu$Gal, is identified.

In summary, the Earth's tide correction is expressed as
\begin{equation}
	\Delta g_{\mathrm{tt}}=k_{\mathrm{op}}\cdot g(\mathrm{Tsoft})+\Delta g_{\mathrm{o}}.
\end{equation}
The standard uncertainty associated with the Earth's tide correction is 0.8\,$\upmu$Gal.

\subsection{Self-attraction correction, $\Delta g_{\mathrm{s}}$}

The mass distribution within the Kibble balance instrument induces a gravitational field, thereby necessitating a self-attraction correction for $g$. The self-attraction effect in a conventional Kibble balance typically exhibits an impact on the order of a few parts in $10^8$. Given the $g$ measurement {relative uncertainty} target of $10^{-9}$, a determination of this effect with a 10\% uncertainty is deemed sufficient. At this precision level, three methods are viable: 1) direct calculation of the effect based on Newton’s law of gravitation, 2) measurement of the relative gravity change using gravimeters, and 3) finite element analysis (FEA) calculations utilizing the analogy between Coulomb's law and Newton’s law of gravitation. For tabletop Kibble balances, direct measurement proves challenging due to spatial constraints. Consequently, FEA calculations are employed in this context.


\begin{figure}[tp!]
	\centering
	\includegraphics[width=0.425\textwidth]{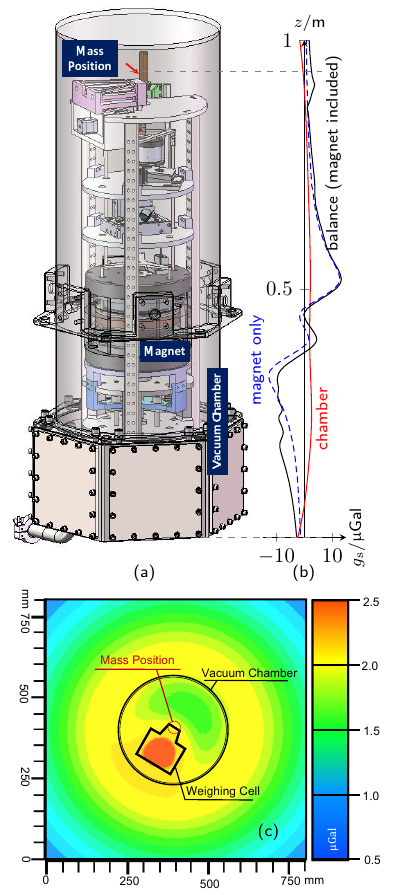}
	\caption{(a) shows the CAD model of the Tsinghua tabletop Kibble balance. (b) present the self-attraction effect of major mechanical pieces as a function of vertical distance. Note that the zero position $z=0$ is set at the bottom of the vacuum chamber. (c) plots the gravitational field map at the weighing horizontal plane.}
	\label{fig:THUKB}
\end{figure}

The self-attraction field generated by a focused mechanical segment is written as
\begin{equation}
	\Delta g_{\mathrm{s}}=G\frac{\rho V}{|r|^3}\overrightarrow{r}\cdot \overrightarrow{k},
	\label{eq:gfield}
\end{equation}
where $G$ is the gravitational constant, $\rho$ and $V$ respectively the density and volume of the segment, $\overrightarrow{r}$ the vector from {the mass position} to the segment mass center, $\overrightarrow{k}$ the unit vector along the vertical direction $z$. 
When the segment is imported into the FEA package, it needs to first change the material into the vacuum and then set a constant charge density numerically $\rho'=\rho$. Finally, the vertical electrical field at {the mass position} $E_z$ can be solved by FEA following
\begin{equation}
	E_z=\frac{1}{4\pi\epsilon_0} \frac{\rho' V}{|r|^3}\overrightarrow{r}\cdot \overrightarrow{k},
	\label{eq:efield}
\end{equation}
where $\epsilon_0$ is the vacuum permittivity.
Comparing (\ref{eq:gfield}) and (\ref{eq:efield}), the self-attraction is determined as 
\begin{equation}
	\Delta g_{\mathrm{s}}=4\pi\epsilon_0GE_z.
\end{equation}

The CAD model of Tsinghua tabletop Kibble balance is shown in Fig.\ref{fig:THUKB}(a). In the calculation, only a few major components, including the vacuum chamber, the balance major framework, and the magnet, are considered. The self-attractive field, $g_s$, along the vertical direction, is shown in Fig.~\ref{fig:THUKB}(b). The reference position, $z=0$, is chosen at the bottom of the vacuum chamber and {the mass position} is at $z_\mathrm{w}=0.918$\,m. Note that $z_\mathrm{w}$ contains half of the height of a stainless steel mass (64\,mm/2=32\,mm). It can be seen that {the relative uncertainty caused by} the self-attraction at {the mass position} is well below $1\times10^{-8}$. The contribution of the chamber and the balance (including the magnet) is respectively -0.2\,$\upmu$Gal and 2.2\,$\upmu$Gal. In Fig.~\ref{fig:THUKB}(c), the self-attraction distribution at the horizontal plane of {the mass position} (stainless steel mass) is mapped. The self-attraction value varies slightly inside the vacuum chamber, mainly due to the near-field mass distribution. 

A specific feature of the Tsinghua Kibble balance is to move the magnet instead of the coil during the velocity measurement. The movement of the magnet may reshape the self-attraction curve along the vertical. However, the gravitational field change is about 0.3\,$\upmu$Gal over a 10\,mm measurement range, which compared to the measurement uncertainty target is negligible.   

The uncertainty of self-attraction correction mainly depends on the material density and segment dimensions, since the calculation accuracy is, in general, well below 10\%, a standard uncertainty of 0.5\,$\upmu$Gal is assigned. 

\subsection{Final result of $g$ determination}

\begin{table*}[tp!]
	\centering
	\caption{Summary of the $g$ determination}
	\label{tab:gresult}
	\begin{tabular}{clrc}
		\hline
		&  method	            & value/$\upmu$Gal   & uncertainty ($k=2$)/$\upmu$Gal\\
		\hline
		$g_{\rm{r}}$	      & key comparison        & 980\,122\,922.8        & 2.0\\
		$g_{\rm{0}}-g_{\rm{r}}$       & tide correction       & 16\,658.4	         & 4.2\\
		& polynomial fit ($O=6$)        & 16\,657.6            & 2.4\\
		& weighted mean         & 16\,657.7            & 2.0\\
		$g_{\rm{0}}$           & $g_{\rm{r}}+(g_{\rm{0}}-g_{\rm{r}})$       & 980\,139\,580.5        & 2.8\\
		$\Delta g_\mathrm{HGG}$ &gravity mapping($z=0$)		& 1.2                & 3.0\\
		&gravity mapping($z=0.3$\,m)		&-0.7                & 5.4\\
		&gravity mapping($z=0.6$\,m)		&3.2                & 1.8\\
		&weighted mean		                &2.4               & 1.4\\
		$\Delta g_\mathrm{VGG}$ &gravity mapping & -296.2($H_m$/m-0.212)            & 3.6 \\
		$g_{\rm{m}}$			&$g_0+\Delta g_\mathrm{HGG}+\Delta g_\mathrm{VGG}$ &  -    &4.8\\
		$\Delta g_{\rm{ta}}$			&US Standard Atmosphere 1976  &     -      &1.0\\
		$\Delta g_{\rm{tp}}$			&Recommended formula  &     -      &0.2\\
		$\Delta g_{\rm{tt}}$			&Tsoft estimation  &       -    &1.6\\
		$\Delta g_{\rm{t}}$          &$\Delta g_{\rm{ta}}+\Delta g_{\rm{tp}}+\Delta g_{\rm{tt}}$ & - &2.0\\
		$\Delta g_{\rm{s}}$			&FEA, inverse square laws &    2.0      &1.0\\
		$g$			      & $g_{\rm{m}}+\Delta g_{\rm{t}}+\Delta g_{\rm{s}}$   &   -    &5.4 \\
		\hline
	\end{tabular}
	
\end{table*}

\begin{figure}[tp!]
	\centering
	\includegraphics[width=0.5\textwidth]{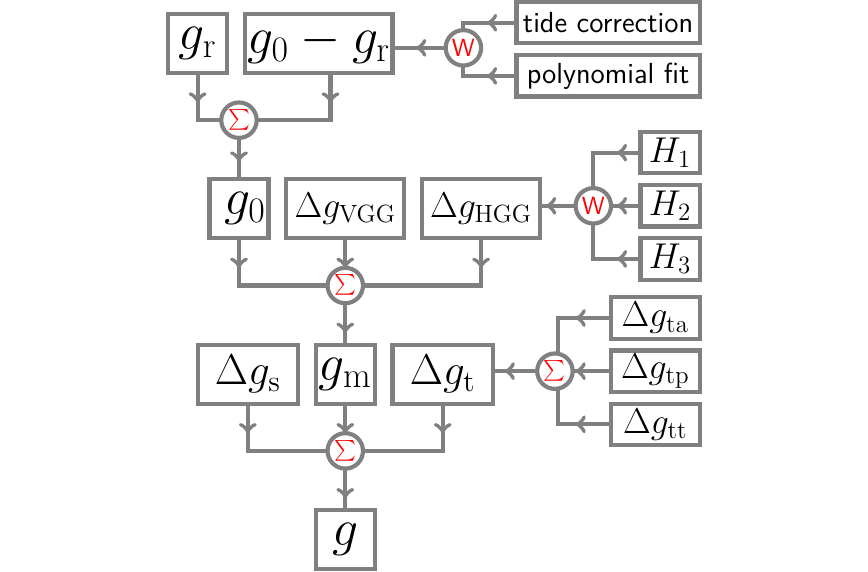}
	\caption{{A cause-and-effect graph of the value and the uncertainty components up to the determination of the final $g$ value. Here $\Sigma$ denotes that the output quantity equals the sum of the inputs, while W means the weighted mean.}}
	\label{fig:figoftable2}
\end{figure}

With all the measurement and analysis, we can finally summarize the $g$ determination result. The measurement uncertainty budget is presented in Tab. \ref{tab:gresult}, {including the weighted mean of uncertainties from different methods, and merging different uncertainty components.} 
{A cause-and-effect graph of the value and the uncertainty components up to the determination of the final $g$ value is also given in Fig.\ref{fig:figoftable2}.}
The total $g$ measurement uncertainty obtained is 5.4\,$\upmu$Gal ($k=2$).

{Specifically, the final $g$ value is the sum of $g_{\rm{m}}$, $\Delta g_{\rm{t}}$ and $\Delta g_{\rm{s}}$, therefore, the total $g$ measurement uncertainty is obtained by the square law using the uncertainties of $g_{\rm{m}}$, $\Delta g_{\rm{t}}$ and $\Delta g_{\rm{s}}$, i.e. $u_{g}=\sqrt{(u_{g_{\rm{m}}})^2+(u_{\Delta g_{\rm{t}}})^2+(u_{\Delta g_{\rm{s}}})^2}.$ The uncertainties of $g_{\rm{0}}$, $g_{\rm{m}}$, and $\Delta g_{\rm{t}}$ can be similarly determined. For ($g_{\rm{0}}-g_{\rm{r}}$) and $\Delta g_{\rm{HGG}}$, the average value and the uncertainty are obtained by the weighted mean of results corresponding to the different measurement methods. The weight for each method, $w_i$, is set as $w_i=1/u_i^2$, where $u_i$ is the measurement uncertainty for the $i$-th method.}

\section{Conclusion}

This paper presents the determination of the local gravitational acceleration, $g$, for the Tsinghua tabletop Kibble balance with a targeted {relative} uncertainty of $10^{-9}$. The gravity transfer between an international gravity comparison site at NIM Changping campus and the Kibble balance laboratory at Tsinghua University has been carried out, alternately realized by using  a 
relative gravimeter CG6. Beyond the conventional tide correction, a polynomial fitting method was introduced for the blind transfer of absolute gravitational acceleration. The results obtained through this approach exhibit great convenience with a good agreement with values derived from the tide correction. Furthermore, the investigation involved the extraction of horizontal and vertical gravity gradients by mapping the gravity distribution at various heights, providing valuable insights into the spatial variations of gravitational acceleration. To enhance the precision of the Kibble balance experiment, comprehensive modeling of time-varying effects and the self-attraction effect was conducted.
The culmination of these efforts resulted in the achievement of the final determination of the gravitational acceleration at {the mass position} for the Tsinghua tabletop Kibble balance experiment. The reported uncertainty stands at 5.4\,$\upmu$Gal ($k=2$), emphasizing a base for future precise realization of masses at the $10^{-8}$ level. 

{Looking ahead, two key tasks are planned: a direct $g$ value check using absolute gravimeters and the vibration measurement. Once the experiment is fully set up, the direct $g$ measurement will be conducted to complement and validate the $g$ transfer method used in this study. Additionally, since the Tsinghua tabletop Kibble balance will be installed on the fourth floor of the building, addressing the effects of mechanical vibrations transmitted through the ground is crucial, particularly along the vertical axis. Vibration sensors will be installed at various locations within the building and inside the experiment to ensure accurate and reliable measurements.}

\section*{Acknowledgement}
The authors would like to thank the Uber driver, Mr. Liu, for the 14-hour non-stop delivery of the CG6 instrument between the Tsinghua University site and the NIM Changping campus site. Thanks also to Zhilan Huang at Tsinghua University for organizing the gravity transfer. Shisong Li would like to thank his BIPM colleagues for the valuable discussions on the $g$ measurement. 


\end{document}